\documentclass{iaus}
\usepackage{graphicx}
\def\teff{$T_{\rm eff}$}
\def\MoH{${\rm [M/H]}$}

\title[Elemental abundances in late-type giants with 3D atmosphere models] {Can we trust elemental abundances derived in late-type giants with the classical 1D stellar atmosphere models?}

\author[A. Ku\v{c}inskas et al.]
{
 A. Ku\v{c}inskas$^{1,2}$,
 V. Dobrovolskas$^2$,
 A. Ivanauskas$^{2,1}$,
 H.-G. Ludwig$^3$,
 E. Caffau$^3$,
 K. Bla\v{z}evi\v{c}ius$^4$,
 J. Klevas$^{5,1}$
 \and D. Prakapavi\v{c}ius$^6$
}

\affiliation
{
 $^1$Institute of Theoretical Physics and Astronomy, Go\v{s}tauto 12, Vilnius LT-01108, Lithuania \\ email: {\tt ak@itpa.lt}\\[\affilskip]
 $^2$Vilnius University Astronomical Observatory, \v{C}iurlionio 29, Vilnius LT-03100, Lithuania\\[\affilskip]
 $^3$GEPI - CIFIST, Observatoire de Paris-Meudon, 5 place Jules Janssen, \\92195 Meudon Cedex , France\\[\affilskip]
 $^4$Department of Applied Mathematics and Informatics, Vilnius University, Naugarduko 24, LT-03225 Vilnius, Lithuania\\[\affilskip]
 $^5$Department of Physics, Vilnius University, Saul\.{e}tekio 9, LT-10222 Vilnius, Lithuania\\[\affilskip]
 $^6$Department of Physics, The University of Liverpool, Liverpool L69 7ZE, UK
}

\pubyear{2009}
\volume{265}  
\pagerange{119--126}
\jname{Chemical Abundances in the Universe: Connecting First Stars to Planets}
\editors{K. Cunha, M. Spite \& B. Barbuy, eds.}
\begin{document}

\maketitle

\begin{abstract}
We compare the abundances of various chemical species as derived with 3D hydrodynamical and classical 1D stellar atmosphere codes in a late-type giant characterized by \teff$=3640$\,K, $\log g=1.0$, $\MoH=0.0$. For this particular set of atmospheric parameters the 3D--1D abundance differences are generally small for neutral atoms and molecules but they may reach up to 0.3--0.4\,dex in case of ions. The 3D--1D differences generally become increasingly more negative at higher excitation potentials and are typically largest in the optical wavelength range. Their sign can be both positive and negative, and depends on the excitation potential and wavelength of a given spectral line. While our results  obtained with this particular late-type giant model suggest that 1D stellar atmosphere models may be safe to use with neutral atoms and molecules, care should be taken if they are exploited with ions.
\end{abstract}

Current 3-dimensional hydrodynamical codes have taken stellar atmosphere modeling to a new level of realism, making it possible to assess the influence of various non-stationary phenomena (e.g., convection) on the observable properties of various classes of stars. It has been recently demonstrated by \cite[Collet et al. (2007)]{CAT07} that in the domain of late-type giants, which are important tracers of intermediate age and old stellar populations, the use of 1D classical and 3D hydrodynamical stellar atmosphere models may result in substantially different elemental abundances, especially at  $\MoH \lesssim-2.0$. However, since \cite[Collet et al. (2007)]{CAT07} have studied only stars located close to the base of RGB it is not clear whether similar conclusions would apply at lower effective temperatures and gravities.

\begin{figure}[t]
\begin{center}
 \includegraphics[width=4.6in]{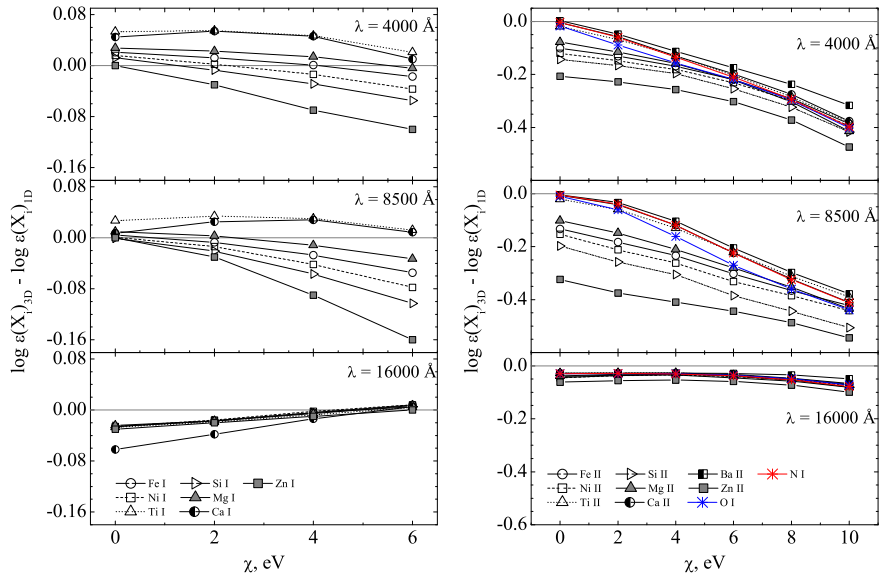}
 \caption{The 3D--1D abundance differences plotted versus excitation potential, $\chi$, and shown at several wavelengths for neutral (left panel) and ionized species (plus N\,I and O\,I, right panel) in a late-type giant characterized by \teff$=3640$\,K, $\log g=1.0$, $\MoH=0.0$.}
   \label{fig1}
\end{center}
\end{figure}

In this work we used a model of a considerably cooler late-type giant (\teff$=3640$\,K, $\log g=1.0$, $\MoH=0.0$) to assess the 3D--1D abundance differences at low effective temperatures. For this purpose we synthesized a number of artificial spectral lines at different wavelengths and with different excitation potentials, corresponding to various neutral atoms, ions and molecules. 3D and 1D stellar atmosphere models used in the spectral line synthesis were calculated with the {\tt CO$^5$BOLD} and {\tt LHD} stellar atmosphere codes using identical atmospheric parameters (\cite[Freytag et al. 2002]{F02}; \cite[Freytag et al. 2003]{F03}; \cite[Wedemeyer et al. 2004]{W02}). The {\tt CO$^5$BOLD} simulation was run on a Cartesian grid of 150x150x151 grid points (xyz, respectively). Both models shared identical equation of state and opacities.

Generally, at a fixed abundance $\epsilon(X_{\rm i})$ the strength of a spectral line corresponding to chemical element $X_{\rm i}$ will be different when calculated with 3D hydrodynamical and classical 1D models. To evaluate the size of these differences we derived abundances of the element $X_{\rm i}$, $\log \epsilon(X_{\rm i})_{3D}$ and $\log \epsilon(X_{\rm i})_{1D}$, which produce the same equivalent width of a given spectral line with the 3D and 1D models. The 3D--1D abundance difference was then defined as $\log \epsilon(X_{\rm i})_{3D}$--$\log \epsilon(X_{\rm i})_{1D}$. It was always derived for the weakest spectral lines ($<10$\,m{\AA}), to minimize effects associated with the choice of microturbulent velocity in the 1D model which affects the equivalent widths of stronger lines.

We find that for the neutral atoms the 3D--1D abundance differences are small, typically $-0.1\dots+0.05$\,dex (Fig.~1). The differences tend to become positive with lower excitation potentials, with the exception of near-infrared spectral lines which show a weak opposite trend. Similarly, the differences in case of molecules are in the range $-0.1\dots+0.1$\,dex. They may be both positive and negative and for lines of the same molecule may alternate sign at different wavelengths. The largest 3D--1D differences are seen in case of ions where they may reach $-0.4$\,dex (Fig.~1). Here, except for the near-infrared lines, the 3D--1D differences grow increasingly more negative at higher excitation potentials. Qualitatively, the latter finding is similar to what was obtained by \cite[Steffen \& Holweger (2002)]{SH02}, with the exception that the 3D--1D differences derived in our study are significantly larger.  Noteworthy, abundance differences for all lines are small at 16000{\AA}, typically $<0.05$\,dex.

\end{document}